\documentclass[%
 reprint,
 amsmath,amssymb,
 aps,
 superscriptaddress,
 prd,
]{revtex4-2}

\usepackage[colorlinks,
           bookmarks=true,
           linkcolor=blue,
           urlcolor=blue, 
            anchorcolor=black,
            citecolor=blue
            ]{hyperref}

\usepackage{multirow}
\usepackage{graphicx}
\usepackage{dcolumn}
\usepackage{bm}
\usepackage{float}
\usepackage{exscale}
\usepackage{relsize}
\usepackage{amsmath}

\begin{document}

\title{X(2370) glueball-like particle productions in $e^+e^-$ collisions at the BESIII energy and in pp collisions 
at the LHC energy with PACIAE model}
 
\author{Jian Cao}
\affiliation{School of Physics and Information Technology, Shaanxi Normal University, Xi'an 710119, China}
\author{Zhi-Lei She}     
\email{shezhilei@cug.edu.cn}
\affiliation{Wuhan Textile University, Wuhan 430200, China}
\author{Jin-Peng Zhang}
\affiliation{School of Physics and Information Technology, Shaanxi Normal University, Xi'an 710119, China}
\author{Jia-Hao Shi}
\affiliation{School of Physics and Information Technology, Shaanxi Normal University, Xi'an 710119, China}
\author{Zhi-Ying Qin}
\affiliation{School of Physics and Information Technology, Shaanxi Normal University, Xi'an 710119, China}
\author{Wen-Chao Zhang}
\email{wenchao.zhang@snnu.edu.cn}
\affiliation{School of Physics and Information Technology, Shaanxi Normal University, Xi'an 710119, China}
\author{Hua Zheng}
\affiliation{School of Physics and Information Technology, Shaanxi Normal University, Xi'an 710119, China}
\author{An-Ke Lei}
\affiliation{Key Laboratory of Quark and Lepton Physics (MOE) and Institute of
            Particle Physics, Central China Normal University, Wuhan 430079,
            China}
\author{Dai-Mei Zhou}
\email{zhoudm@mail.ccnu.edu.cn}
\affiliation{Key Laboratory of Quark and Lepton Physics (MOE) and Institute of
            Particle Physics, Central China Normal University, Wuhan 430079,
            China}
\author{Yu-Liang Yan}
\affiliation{China Institute of Atomic Energy, P. O. Box 275 (10), Beijing
            102413, China}
\author{Ben-Hao Sa}
\email{sabhliuym35@qq.com} 
\affiliation{China Institute of Atomic Energy, P. O. Box 275 (10), Beijing
            102413, China}           
\date{\today}

\begin{abstract}
Inspired by the BESIII newest observation of X(2370) glueball-like particle, we search its productions in both $e^+e^-$ collisions at $\sqrt{s}=$ 4.95 GeV and proton-proton (pp) collisions at $\sqrt{s}=$ 13 TeV with a parton and hadron cascade model 
PACIAE. In this model, the 
final partonic state (FPS) and the final hadronic state (FHS) are consecutively 
simulated and recorded. The X(2370) glueball- or tetraquark-state is then, 
respectively, recombined by two gluons or four quarks $ss\bar{s}\bar{s}$ in the 
FPS using the quantum statistical mechanics inspired dynamically constrained 
phase-space coalescence (DCPC) model. The X(2370) molecular-state is 
recombined by the baryon-antibaryon of $\Lambda$-$\bar{\Lambda}$ or 
$\Sigma$-$\bar{\Sigma}$, or by three mesons of $\pi^+\pi^{-}\eta'$,  
 $K^+K^-\eta'$, or $K_S^0K_S^0\eta'$ in the FHS using DCPC model. In both $e^+e^-$ and pp collisions, significant discrepancies in the yields, the transverse momentum spectra and the rapidity 
distributions among the X(2370) glueball-, tetraquark-, and 
molecular-state are observed. These discrepancies are proposed as valuable criteria identifying the X(2370) different states from each other. Our results not only support the BESIII observation of glueball-like particle $\rm X(2370)$ production in $e^+e^-$ collisions, but also serve as a prediction for the $\rm X(2370)$ production in pp collisions. We strongly suggest the experimental measurement of the X(2370) glueball-like particle production in pp collisions at the LHC energies.

\end{abstract}

\maketitle

\section{\label{sec:intro}Introduction}
The constituent quark model \cite{quark_model_1, quark_model_2} has been the basic framework within which the conventional mesons and baryons could be understood. The non-Abelian property of quantum chromodynamics (QCD) permits the existence of new types of hadrons, such as the glueballs, hybrid states, hadronic molecular states and multiquark states \cite{new_hadron_1,new_hadron_2,new_hadron_3}. In particular, glueballs are unique particles which are bound states of gluons on their own, without any quarks involved. The Lattice QCD (LQCD) in the quenched approximation predicts that in the ground state the masses of the  scalar, tensor and pseudo-scalar two-gluon glueballs are, respectively, around 1.5-1.7 $\rm GeV/c^{2}$, 2.3-2.4 $\rm GeV/c^{2}$ and 2.3-2.6 $\rm GeV/c^{2}$ \cite{glueball_1,glueball_2,glueball_3,glueball_4,glueball_5}. The hunting of glueballs is one of the important goals in hadron physics. A large number of experimental studies have been conducted in order to confirm their existence over the past four decades, mostly in  radiative decays from $J/\psi$ in $e^+e^-$ collisions \cite{glueball_hunt_1, glueball_hunt_2, glueball_hunt_3}. There are several glueball candidates, such as the scalar mesons $f_0(1500)$ and $f_0(1710)$, the tensor meson $f_2(2340)$, and the pseudoscalar meson X(2370) \cite{glueball_hunt_3}. Among them, the X(2370) is a good candidate for the $0^{-+}$ glueball, as its mass, production and decay properties are consistent with the LQCD prediction \cite{X_2370_1}. 

The X(2370) was first observed in the $\pi^+ \pi^- \eta'$ invariant mass distribution of the $J/\psi\rightarrow \gamma \pi^+ \pi^- \eta'$ decay by the BESIII collaboration \cite{X_2370_1}. Later, it was confirmed by the collaboration in the combined measurement of $J/\psi\rightarrow  \gamma K^+ K^- \eta'$ and $J/\psi\rightarrow  \gamma K_S^0 K_S^0 \eta'$ \cite{X_2370_2}. Recently, the spin-parity of the X(2370) was determined to be $0^{-+}$ for the first time in the decay $J/\psi\rightarrow  \gamma K_S^0 K_S^0 \eta'$ \cite{X_2370_3}. The experimental observation stimulated a number of theoretical interpretations for the  X(2370), such as the fourth radial excitation of $\eta/\eta'$ \cite{X_2370_inter_1,X_2370_inter_2}, the $P$-wave $ss\bar{s}\bar{s}$ tetraquark state \cite{X_2370_inter_3}, the light baryonium states $\Lambda$-$\bar{\Lambda}$ and $\Sigma$-$\bar{\Sigma}$ \cite{X_2370_inter_4}. One of the intriguing explanations is that the X(2370) is a pseudoscalar glueball \cite{glueball_5}.  There are two possible compositions for the pseudoscalar glueball: two- or three-gluons \cite{glueball_config}. The  experimental result  at BESIII
 is in favor of  the X(2370) as the two-gluon glueball  structure \cite{X_2370_3}.
The X(2370) could not be a pseudoscalar glueball composed of three  gluons, as the mass prediction from the quenched LQCD for this configuration is around 3.4-3.6 GeV/c$^2$ \cite{glueball_mass_predict_1, glueball_mass_predict_2}, which is much heavier than the mass of the X(2370) measured. 

Apart from the radiative $J/\psi$ decay in $e^+e^-$ collisions, the proton-proton (pp) collisions are also theoretically suggested to search for glueballs \cite{glueball_mass_predict_1}.  In this paper, we carry out the investigation of the X(2370) productions in both $e^+e^-$ collisions at the center-of-mass energy $\sqrt{s}=$ 4.95 GeV and pp collisions at  $\sqrt{s}$ =  
13 TeV with a parton and hadron cascade model PACIAE \cite{paciae_3}. In the 
model, the final partonic state (FPS) and the final hadronic  state (FHS) are 
consecutively simulated and recorded. There are three scenarios considered for 
the configuration of the X(2370): the glueball-, tetraquark- and 
molecular-state. The glueball- and tetraquark-state are, respectively, 
produced by coalescing with two gluons and four quarks of $ss\bar{s}\bar{s}$ in the FPS using the quantum statistical mechanics inspired dynamically constrained phase-space coalescence (DCPC) model \cite{DCPC}. The molecular-state is generated by recombining the baryon-antibaryon ($B$-$\bar{B}$) of $\Lambda$-$\bar{\Lambda}$ or $\Sigma$-$\bar{\Sigma}$, or three mesons of $\pi^+\pi^{-}\eta'$, $K^+K^-\eta'$ or $K_S^0K_S^0\eta'$ in the FHS. The resulted X(2370) yields, rapidity ($y$) distributions,  transverse momentum ($p_{\rm T}$) single-differential distributions, as well as $p_{\rm T}$ and $y$ double-differential distributions in both $e^+e^-$ and pp collisions show significant discrepancies between the X(2370) different states. These discrepancies are proposed as valuable criteria identifying the X(2370) different states from each other. Our results not only support the BESIII observation of glueball-like particle $\rm X(2370)$ production in $e^+e^-$ collisions, but also serve as a prediction for the $\rm X(2370)$ production in pp collisions. We strongly suggest the experimental measurement of the X(2370) glueball-like particle production in pp collisions at the LHC energies.

\begin{figure*}[]
\includegraphics[scale=0.43]{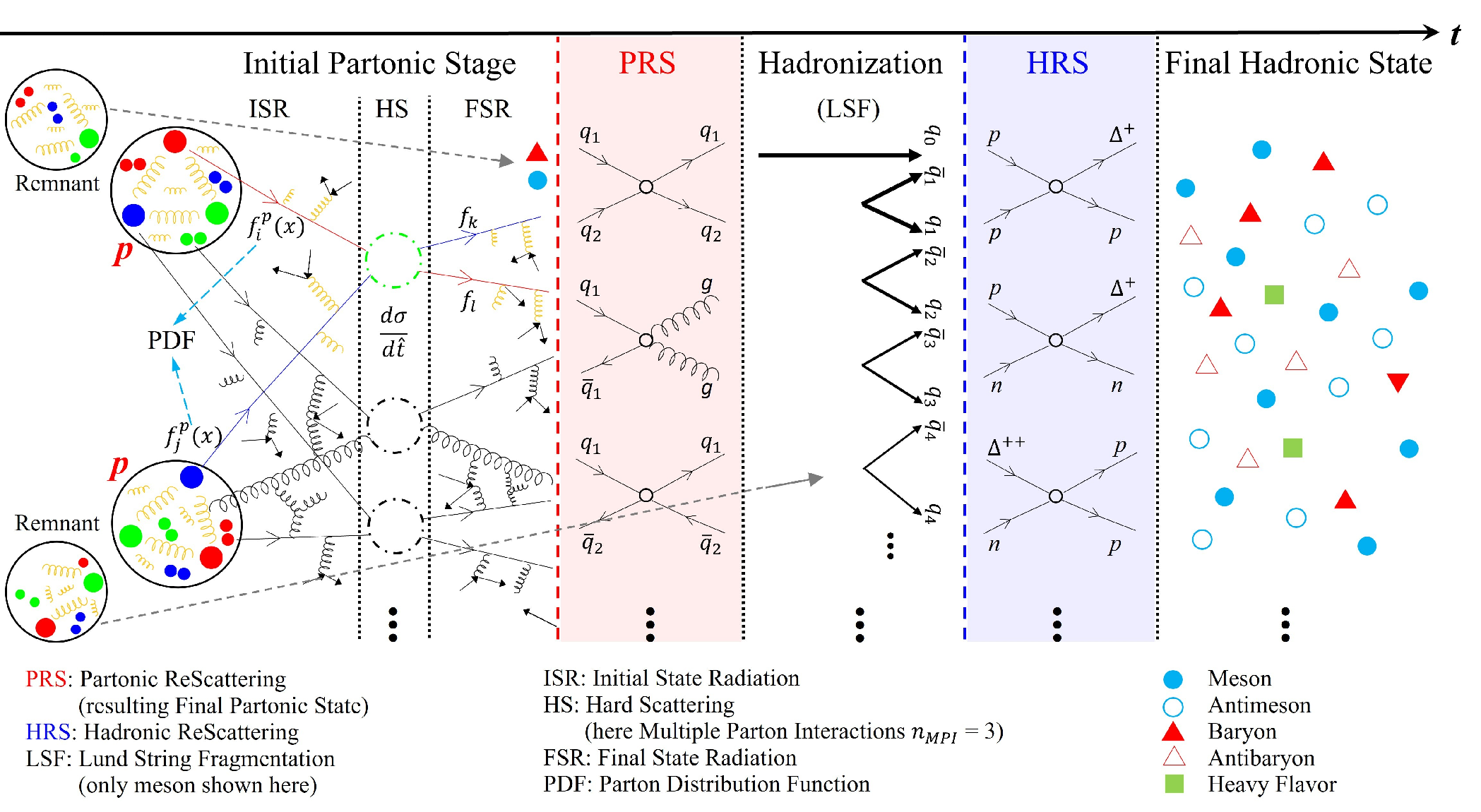}
\caption{\label{fig:physical_routine}  A sketch of the physical routines in high energy pp collisions.}
\end{figure*}

\section{\label{sec:model}The PACIAE and DCPC models}
The PACIAE model is designed for the relativistic elementary collisions and 
nuclear collisions \cite{PACIAE_2}. It is based on PYTHIA 6.4 code \cite{pythia_6} 
but extended considering the partonic rescattering before hadronization and 
the hadronic rescattering after hadronization. It has been successfully 
applied in describing the hadron yields, $p_{\rm T}$ spectra and $y$ 
distributions \cite{paciae_3, PACIAE_2, Jpsi_production}, the strangeness 
enhancement\cite{strangeness_enhance_1,strangeness_enhance_2}, the nuclear 
modification factor \cite{RAA_1, RAA_2}, the elliptic flow \cite{v2_1, v2_2}, 
etc., in the high-energy particle and nuclear collisions. In this work, the 
latest version of PACIAE 3.0 \cite{paciae_3} is used to simulate $e^+e^-$ collisions at $\sqrt{s}=$ 4.95 GeV  and pp collisions  at $\sqrt{s}=$ 13 TeV. In PACIAE model, an $e^+e^-$ or a pp collision is developed from the 
initial parton stage, to the parton rescattering stage, the hadronization, and the hadron rescattering stage. As an example, figure \ref{fig:physical_routine} shows a 
sketch of the physical routines in a high-energy pp collision.

\begin{figure}[h]
\includegraphics[scale=0.43]{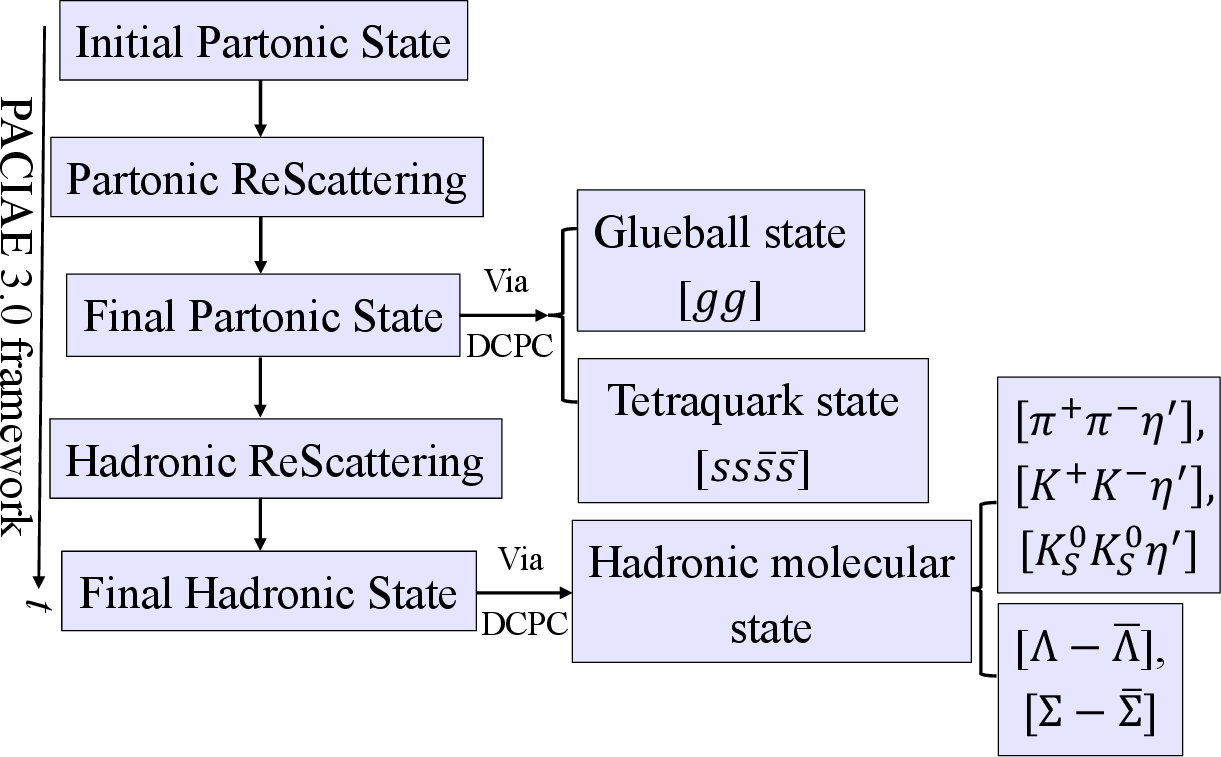}
\caption{\label{fig:glueball_sketch}  A sketch of the X(2370) glueball-, 
tetraquark- and molecular-state productions in $e^+e^-$ collisions at $\sqrt{s}$= 4.95 GeV or in pp collisions at $\sqrt{s}$= 13 TeV  with the PACIAE+DCPC model.}
\end{figure}

In the first stage, both the  $e^+e^-$ and the pp collision is executed by PYTHIA \cite{pythia_6} 
with temporarily switching off the string fragmentation. Thus an initial 
partonic state is available after the parton-parton hard scattering, the 
associated initial- and final-state QCD radiations, the strings broken down 
and the diquarks (anti-diquarks) splitted up. This partonic matter then 
undergoes parton rescatterings, where the leading order (LO) pQCD 
parton-parton scattering cross sections \cite{cs_1, cs_2} are employed. The 
final partonic state is generated after partonic rescattering. It consists of 
numerous quarks, antiquarks and gluons with their four-coordinate and 
four-momentum. In the hadronization stage, the partonic matter is converted 
into hadrons by the string fragmentation scheme \cite{pythia_6}. The followed 
is then the hadronic rescattering resulting in a final hadronic state. It is 
composed of abundant hadrons with their four-coordinate and four-momentum. A 
sketch of the aforementioned  processes in $e^+e^-$ or pp collisions is presented in the left part 
of Fig. \ref{fig:glueball_sketch}.

The DCPC model was proposed by us to study the light nuclei production 
in pp collisions at the LHC energies \cite{DCPC}. It has been successfully applied to 
calculate the yield of the exotic states such as the 
X(3872) \cite{X_3872_1, X_3872_2, X_3872_3}, $Z_c^{\pm}(3900)$ \cite{zc_3900}, 
$P_c(4312)$, $P_c(4440)$, $P_c(4457)$ \cite{pc_states}, and 
$\Omega_c^0$ \cite{Omega_c} after the transport model simulation. In DCPC 
model, the yield of $N$-particle cluster is estimated according to the quantum 
statistical mechanics \cite{DCPC,book1,book2} by 
\begin{equation}
Y_{\rm cluster}=\int\cdots\int_{E_1\le E\le E_2}\frac{d\vec{q}_{1}d\vec{p}_{1}\cdots d\vec{q}_{N}d\vec{p}_{N}}{h^{3N}}, \label{eq:yield_of_N_cluster}
\end{equation}
where $E_1$ and $E_2$ are the cluster's lower and upper energy 
thresholds, $\vec {q}_{i}$ and $\vec {p}_{i}$ are the $i$th particle's 
three-coordinate and three-momentum, respectively. Therefore, the yield 
of the X(2370) two-gluon glueball-state, for instance, reads  
\begin{equation}
Y_{\rm glueball}=\frac{1}{2!}\int\int_{E_1\le E\le E_2}\delta_{12}\frac{d\vec{q}_{1}d\vec{p}_{1} d\vec{q}_{2}d\vec{p}_{2}}{h^{6}}, \label{eq:yield_of_glueball}
\end{equation}
where the factor $1/2!$ is introduced as gluons are identical particles. In the above 
equation $\delta_{12}$ is expressed as 
\begin{equation}
\delta_{12}=
\begin{cases}
1,& \textrm {if} \ 1\equiv g,\ 2\equiv g,\ R_i\leq  R_0,\ \textrm {and} \atop  m_{\rm 0}-\Delta m \leq m_{\rm inv}\leq m_{\rm 0}+\Delta m   \\ 
0,&  \ \ \ \ \ \ \ \ \ \rm otherwise
\end{cases},\label{eq:delta}
 \end{equation}
where $m_0$ refers to the mass of the X(2370),  
$\Delta m$ is the mass uncertainty (a free parameter) which is estimated as 
the half decay width of the X(2370) \cite{X_2370_3}. $R_0$ and $R_i$ are, 
respectively, the radius of the glueball (a free parameter) and the relative 
distance between the component particle $i$ and the center of mass of the 
glueball. $R_0$ is set to be 1 fm as gluons are point-like particles. The 
invariant mass, $m_{\rm inv}$, is calculated as 
$m_{\rm inv}=\sqrt{(E_1+E_2)^2-(\vec p_1+\vec p_2 )^2}$, where $E_{1,2}$ and 
$\vec p_{1,2}$ are, respectively, the energy and momentum of the selected 
gluons 1 and 2. 

The yield of the X(2370) tetraquark- or molecular-state can be evaluated 
in a similar way with different parameters. The parameters of mass uncertainty 
and radius are given in Table \ref{tab:X_2370_para}.

\begin{table}[h]
\caption{The parameters of mass uncertainty and radius for the X(2370) 
glueball-, tentraquark- and molecular-state.}\label{tab:X_2370_para} 
\begin{ruledtabular}
\begin{tabular}{ccccc}
\multirow{2}{*}{} & \multirow{2}{*}{glueball} & \multirow{2}{*}{tetraquark} & \multicolumn{2}{c}{molecular} \\ 
                  &                    &                    & $B$-$\bar{B}$         & 3 mesons         \\
 \colrule                 
 $\Delta m$                  & \multirow{2}{*}{94}                   & \multirow{2}{*}{94}                  & \multirow{2}{*}{94}          & \multirow{2}{*}{94}          \\
 (MeV/c$^{2}$)                 &                    &                   &          &           \\
 \hline
$R_0$ (fm)                & 1.0                  & 1.0                  & 1.0-2.0 \footnote{The $R_0$  upper bound of  the  $B$-$\bar{B}$ molecular-state is taken as the radius  summation of the baryon and antibaryon.}         & 1.0-2.0  \footnote{The $R_0$  upper bound of  the 3-meson molecular-state is assumed to be the same as that of the  $B$-$\bar{B}$ molecular-state.  }       
\end{tabular}
\end{ruledtabular}
\end{table}

To generate the X(2370) two-gluon glueball-state, a 
component particle (gluon) list based on the parton list in FPS is constructed first. Two loops 
over $i$ and $j$ cycling through all gluons in the list are implemented. If 
$i$ and $j$ are two different gluons and they satisfy the constraints in 
Eq. (\ref{eq:delta}), the combination of these two gluons is then deemed as an 
X(2370) glueball-state. The gluon list is then updated by removing the used 
gluons. A new two-layer cycle is executed on the updated list. Repeat these 
steps until the empty of the list or the rest in the list is unable to 
generate an X(2370) glueball-state. The production of the X(2370) tetraquark-state is done in a simiar way. The procedure to generate the molecular-state 
composed of three mesons is a little bit different. A component particle list of 
$\pi^+$, $\pi^-$, $K^+$, $K^-$, $K_S^0$, and $\eta'$ based on the hadron 
list in the FHS is constructed. Three loops cycling through all component 
particles are implemented. Each combination, if it is $\pi^+ + \pi^- + \eta'$, 
$K^+ + K^- + \eta'$, or $K_S^0 + K_S^0 + \eta'$ and satisfies the 
constraints like those in Eq. (\ref{eq:delta}) without the extra factor 
$1/2!$, is assumed to be an X(2370) molecular-state. The generation of the 
X(2370) molecular-state composed of baryon-antibaryon is performed in a similar way.

\begin{figure*}[]
\includegraphics[scale=1]{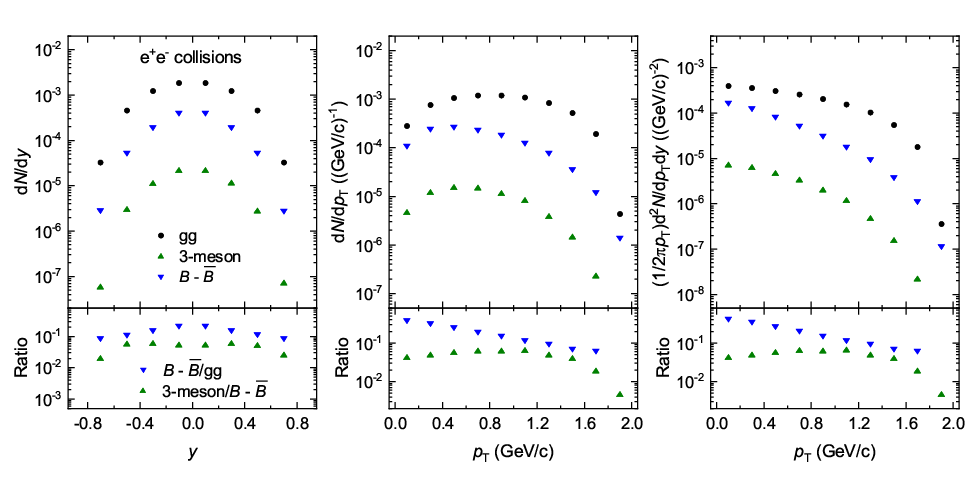}
\caption{\label{fig:e_plus_e_minus}  Upper panles: the simulated $y$ and $p_{\rm T}$ single-differential distributions, as well as $p_{\rm T}$ and $y$ double-differential distributions of
the $\rm X(2370)$ glueball-state, $B$-$\bar{B}$ and 3-meson molecular-state in $e^{+}e^{-}$ collisions at $\sqrt{s}$ = 4.95 GeV. Lower panels: the ratio between two distributions denoted by legend.}
\end{figure*}

\section{\label{sec:results}Results and discussions}
The PACIAE 3.0 model is used to simulate the X(2370) productions in both $e^+e^-$ collisions at $\sqrt{s}$= 4.95 GeV and pp collisions at  $\sqrt{s}$ = 13 TeV. For $e^+e^-$ collisions, default parameters are utilized, as so far there are no experimental yields of pions and kaons available to tune the parameters. For pp collisions, the model parameters are chosen as the default values, except for PARP(31), PARJ(1), PARJ(2) and PARJ(42). 
PARP(31) is a common $K$ factor multiplying the differential cross section 
of hard parton-parton scattering processes. PARJ(1) represents the suppression 
of diquark-antidiquark pair production in string-breaking processes, compared 
with quark-antiquark pair production. PARJ(2) denotes the suppression of 
$s$ quark pair production relative to the $u$ or $d$ pair production. PARJ(42) 
gives the parameter $b$ in Lund fragmentation function. For more details, we refer to Ref. \cite{pythia_6}. These parameters, as listed in Table \ref{tab:PACIAE_parameter}, are fixed by 
fitting to the experimental yields of $\pi^+ + \pi^-$, $K^+ + K^-$, $K_S^0$, 
and $\Lambda + \bar{\Lambda}$ in the mid-rapidity region \cite{ ALICE_yield_3}.
The fitted results of those yields and the corresponding experimental data are 
given in Table \ref{tab:yield_comp}.

\begin{table}[h]
\caption{The PACIAE model parameters of PARP(31), PARJ(1), PARJ(2) and PARJ(42) are
fixed by fitting the $\pi^+ + \pi^-$, $K^+ + K^-$, $K_S^0$, and 
$\Lambda + \bar{\Lambda}$ yields measured in the mid-rapidity region of pp 
collisions at $\sqrt{s}$= 13 TeV.}\label{tab:PACIAE_parameter}
\begin{ruledtabular}
\begin{tabular}{cccc}
  PARP(31) & PARJ(1) & PARJ(2)&PARJ(42)  \\
 \colrule   
  1.0&  0.18 &  0.42&1.35 \\
\end{tabular}
\end{ruledtabular}
\end{table}

\begin{table}[h]
\caption{The PACIAE model simulated yields of $\pi^++\pi^-$, $K^++K^-$, 
$K_S^0$, and $\Lambda+\bar{\Lambda}$ are compared with the ALICE 
data in pp collisions at $\sqrt{s}$= 13 TeV \cite{ALICE_yield_3}.}
\label{tab:yield_comp}
\begin{ruledtabular}
\begin{tabular}{ccccc}
& $\pi^++\pi^-$ & $K^++K^-$ & $K_S^0$ & $\Lambda+\bar{\Lambda}$ \\
 \colrule
Exp. & 4.78$\pm$0.24 & 0.62$\pm$0.03 & 0.32$\pm$0.01 & 0.18$\pm$0.01 \\
PACIAE & 4.83 & 0.65 & 0.32 & 0.18 \\
\end{tabular}
\end{ruledtabular}
\end{table}

\begin{table}[]
\caption{The PACIAE+DCPC model simulated yields of the X(2370) glueball-, 
tetraquark-, and molecular-state in $e^+e^-$ collisions at $\sqrt{s}$= 4.95 GeV and in pp collision at $\sqrt{s}$= 13 TeV.}
\label{tab:X_2370_yield}
\begin{ruledtabular}
\begin{tabular}{ccccc}
  &\multirow{2}{*}{glueball} & \multirow{2}{*}{tetraquark} & \multicolumn{2}{c}{molecular} \\ 
             &   &                    & $B$-$\bar{B}$ \footnote{$B$-$\bar{B}$ refers to the summation over the yields of $\Lambda$-$\bar{\Lambda}$ and $\Sigma$-$\bar{\Sigma}$.}         & 3 mesons \footnote{3-meson refers to the summation over the yields of $\pi^+\pi^{-}\eta'$, $K^+K^-\eta'$ and $K_S^0K_S^0\eta'$.}        \\
     
 $e^+e^-$       &        1.42$\times 10^{-3}$                  & ---                  & 2.63$\times 10^{-4}$          & 1.40$\times 10^{-5}$\\
 pp       &        1.50                  & 2.63$\times 10^{-4}$                  & 4.12$\times 10^{-2}$          & 3.13$\times 10^{-3}$
\end{tabular}
\end{ruledtabular}
\end{table}


Using the PACIAE model, we have generated 800 million  $e^+e^-$ collisions with the default parameters and 100 million pp collision events with the parameters listed in Table \ref{tab:PACIAE_parameter}. As shown in the right 
part of Fig. \ref{fig:glueball_sketch}, the X(2370) glueball- and 
tetraquark-state are generated, respectively, by the recombination of two gluons 
and of four quarks ($ss\bar{s}\bar{s}$) in the FPS using the DCPC model. The 
X(2370) molecular-state is hadronized by the coalescence of baryon-antibaryon 
($\Lambda$-$\bar{\Lambda}$ or $\Sigma$-$\bar{\Sigma}$) or of three mesons 
($\pi^+\pi^{-}\eta'$, $K^+K^-\eta'$ or $K_S^0K_S^0\eta'$) in the FHS. Their yields are given in  Table \ref{tab:X_2370_yield}. The yield of the X(2370) tetraquark-state  in $e^+e^-$ collisions is not available, as at the BESIII energy it is too tiny to be observed. Both the X(2370) glueball- and molecular-state successful generations definitely support the BESIII latest observation of glueball-like particle X(2370) productions in the $e^+e^-$ collisions.

\begin{figure}[h]
\includegraphics[scale=0.33]{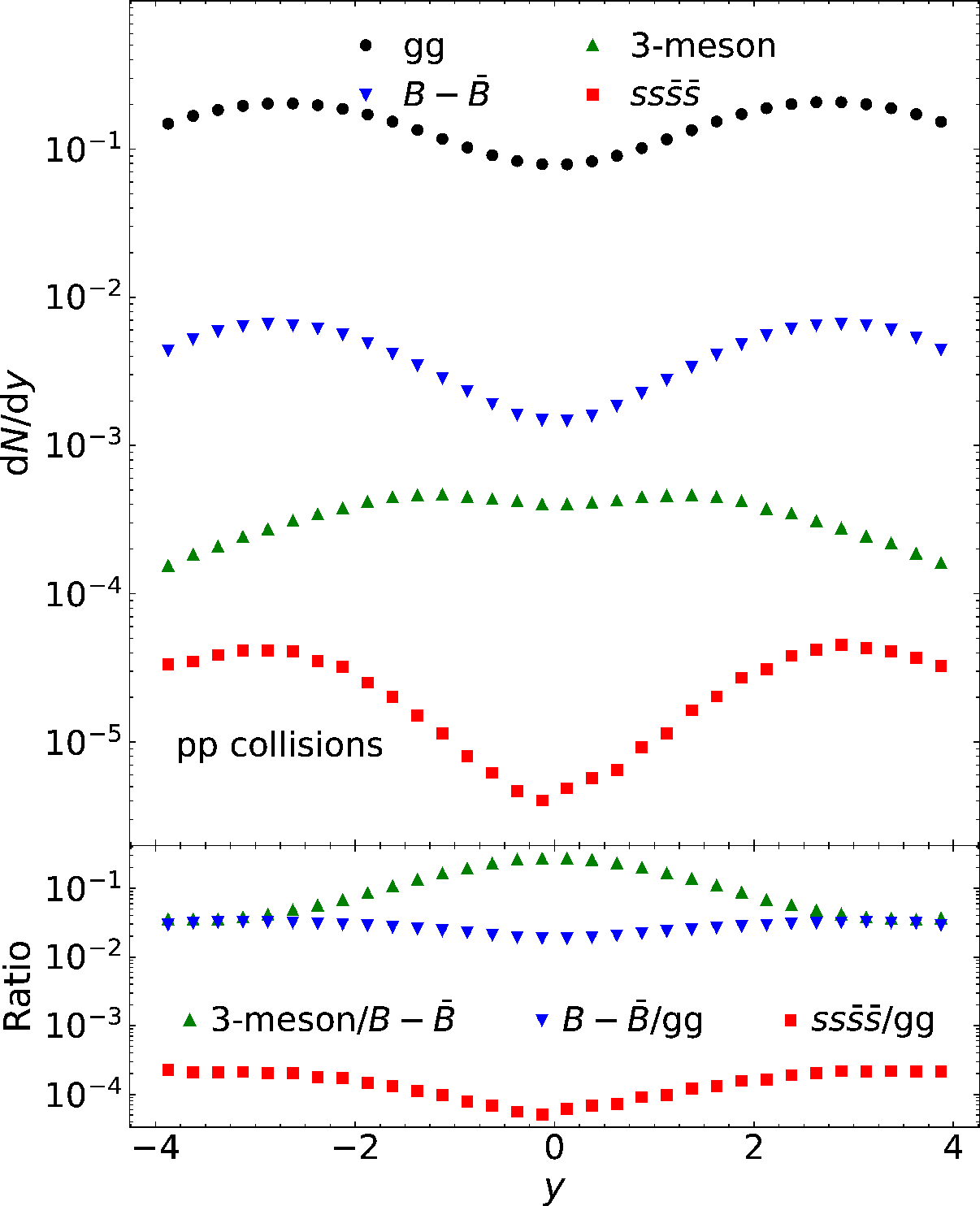}
\caption{\label{fig:single_diff_y_13_TeV}  Upper panel: the simulated $y$ 
single-differential distributions of the X(2370) glueball-state, tetraquark-state, $B$-$\bar{B}$ and  3-meson molecular-state in pp collisions at $\sqrt{s}$= 13 TeV. Lower panel: the ratio 
between two distributions denoted by legend.}
\end{figure}

\begin{figure}[ht]
\includegraphics[scale=0.33]{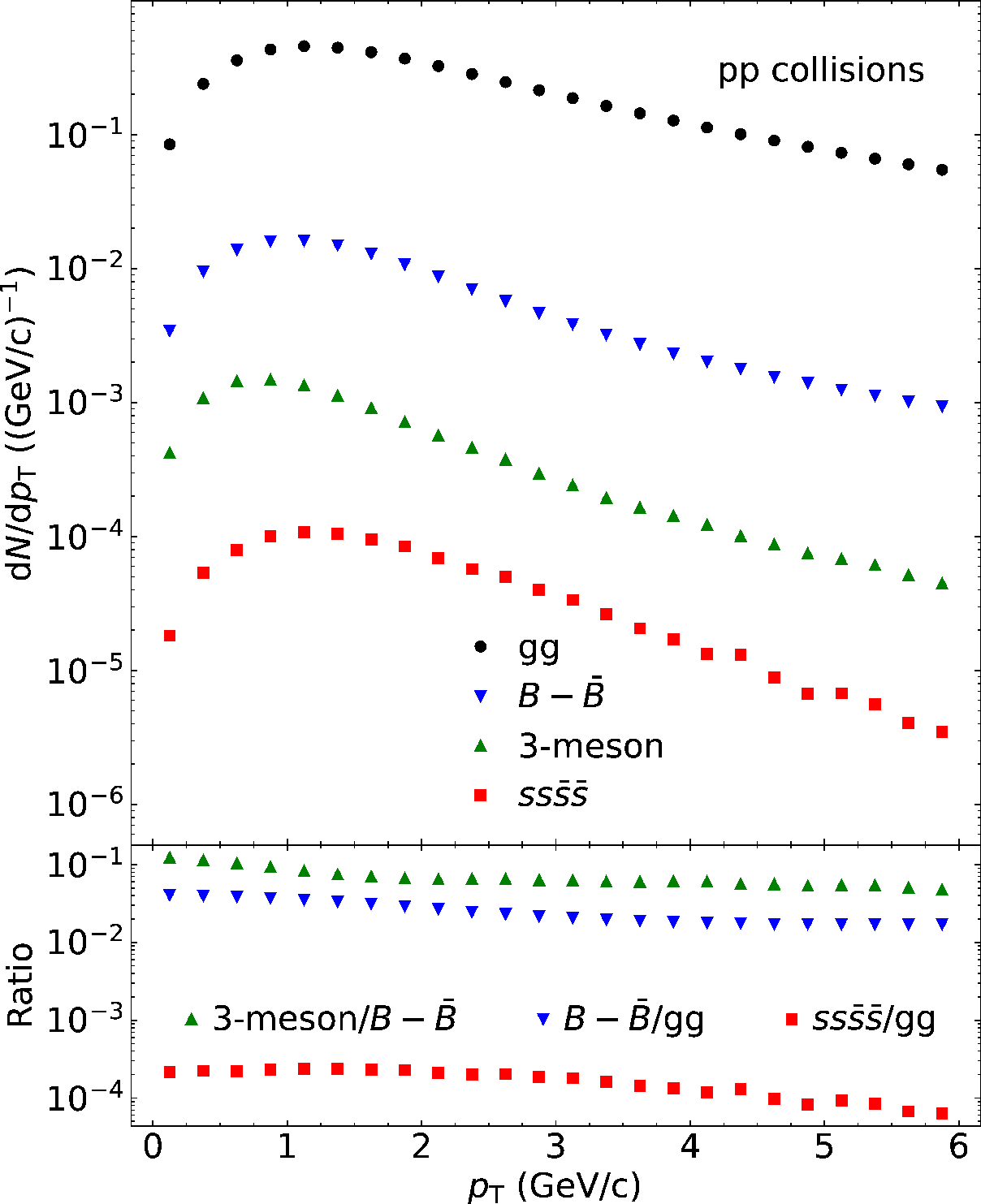}
\caption{\label{fig:single_diff_pt_13_TeV}  The same as in Fig.       \ref{fig:single_diff_y_13_TeV}, but for the $p_{\rm T}$ single-differential distributions.}
\end{figure}

The particle $y$ and $p_{\rm T}$ single-differential distributions, as well as the $p_{\rm T}$ and $y$ double-differential distributions of the X(2370) glueball-state (black circles), $B$-$\bar{B}$ (blue triangles-down) and  3-meson (green triangles-up) molecular-state  in $e^+e^-$ collisions at $\sqrt{s}$ = 4.95 GeV are calculated and shown in Fig. \ref{fig:e_plus_e_minus}. Significant discrepancies between the $\rm X(2370)$ different states are observed in all the above distributions. Thus they can serve as criteria to distinguish
the $\rm X(2370)$ different states from each other. The origin of the different behavior of the three states is linked to the nature of the X(2370), not to the structure of the model, as both our model and a multi-phase transport model (AMPT) \cite{AMPT}  show that the rapidity (and $p_{\rm T}$) distributions of the X(3872) tetraquark- and molecular-state are different in high energy nuclear collisions \cite{X_3872_3, X_3872_4}.

\begin{figure}[ht]
\includegraphics[scale=0.33]{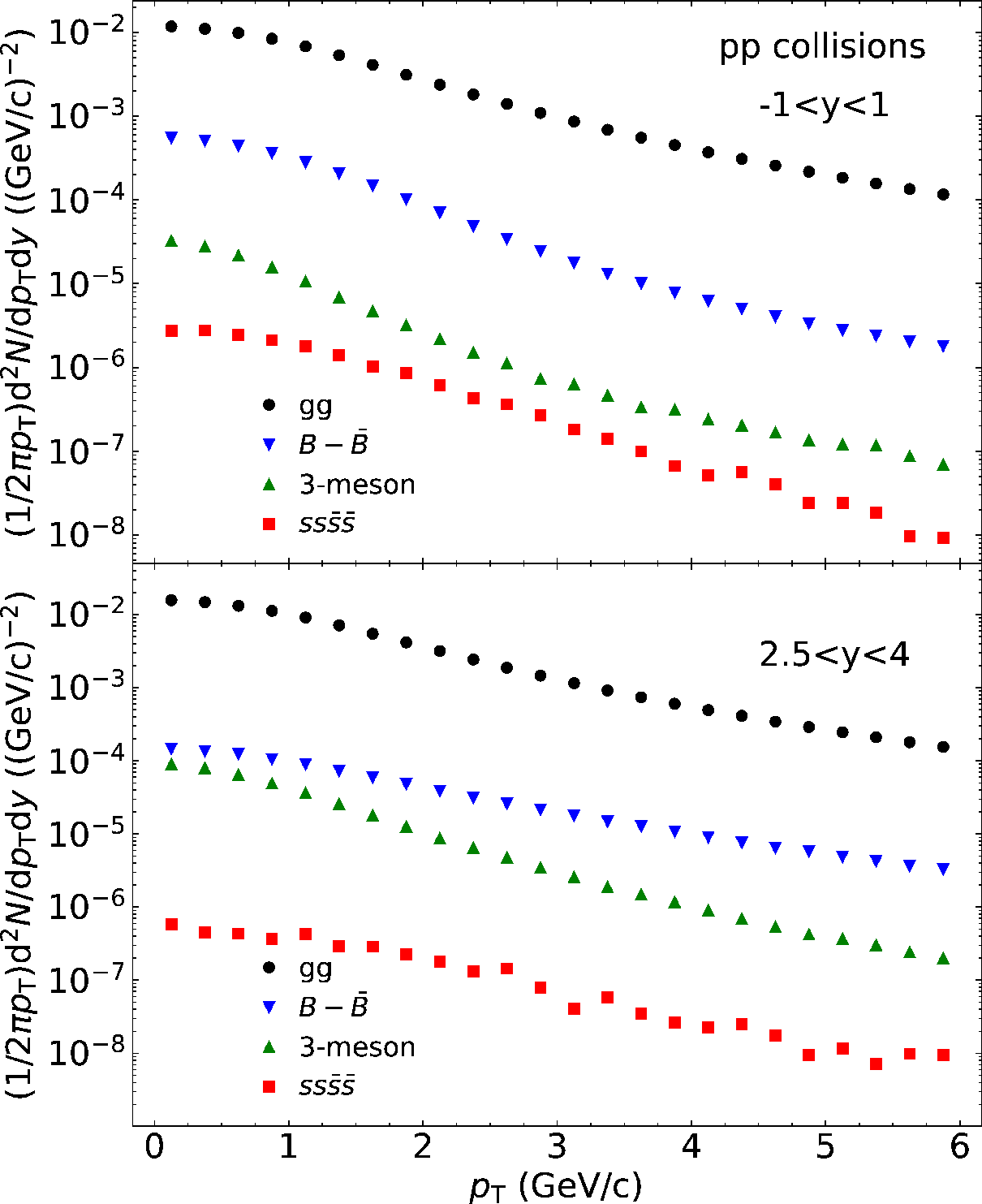}
\caption{\label{fig:cent_fwd_doub_diff_pt_13_TeV}  The simulated $p_{\rm T}$ and $y$ double-differential distributions of the X(2370) glueball-state, tetraquark-state, $B$-$\bar{B}$ and  3-meson molecular-state at mid- and forward-rapidity in pp collisions at $\sqrt{s}$= 13 TeV.}
\end{figure}

The upper panels in Figs. \ref{fig:single_diff_y_13_TeV} and \ref{fig:single_diff_pt_13_TeV} show, respectively, the simulated $y$ and $p_{\rm T}$ single-differential distributions in pp collisions at $\sqrt{s}$= 13 TeV for the X(2370) glueball-state (black circles), 
tetraquark-state (red squares), $B$-$\bar{B}$ (blue triangles-down) and  3-meson (green triangles-up) molecular-state. The upper and lower panels in 
Fig. \ref{fig:cent_fwd_doub_diff_pt_13_TeV} present the simulated $p_{\rm T}$ 
and $y$ double-differential distributions for the X(2370) different states in 
pp collisions at $\sqrt{s}$= 13 TeV in the mid- and forward-rapidity 
regions, respectively. Obvious discrepancies are also observed among the X(2370) different states both in $y$ and $p_{\rm T}$ single-differential distributions, as well as in the $p_{\rm T}$ and $y$ double-differential distributions. Either of them could serve as an effective criterion to identify the X(2370) different states. 



Finally, the same as done in Ref. \cite{X_3872_3}, we explore the temperature evolution in pp collisions. The temperature evolution in $e^+e^-$ collisions is not considered, as the collision energy is low and the partonic matter is not thermalized. With the simulated $u+\bar{u}+d+\bar{d}+g$ in the FPS and $\pi^{+}+\pi^{-}$ in FHS, the Shannon entropies of the partonic matter (PM) and hadronic matter (HM) are, respectively, calculated by \cite{entropy}
\begin{equation}
 \mathcal{H}=-\int_{0}^{\infty}(\mathrm{TMD}/I_{0})\ln[\mathrm{TMD}/I_{0}]dp_{T},
\label{eq:entropy}
\end{equation}	
with TMD and $I_{0}$ being to the $p_{\rm T}$ distribution and its first moment, respectively. The temperatures $T$ of the PM and HM are then determined by \cite{entropy}
\begin{equation}
\mathcal{H}=\frac{m}{m-1}+\ln(\frac{m}{m-1})+\ln(T), 
\label{eq:temperature}
\end{equation}
where $m$ is extrapolated from its energy dependence, $m(\sqrt s)=a_m(\sqrt{s/s_0})^{c_m}$, with $a_m=8.45$, $c_m=-0.082$, $\sqrt{s_0}=$ 1 GeV. The extracted results are  $T_{\rm HM}$ = $T_{\rm molecular}$ = 160 MeV and $T_{\rm QM}$ = $T_{\rm glueball/tetraquark}$ = 232 MeV, which are independent of the DCPC model. The temperature of the HM in pp collisions at $\sqrt{s}$= 13 TeV (160 MeV) is lower than that in pp collisions at $\sqrt{s}$= 2.76 TeV (180 MeV) \cite{X_3872_3}. This is due to the different hadronization mechanisms utilized. In this work, the string fragmentation scheme \cite{pythia_6} is applied while in Ref. \cite{X_3872_3}  the coalescence model \cite{paciae_3} is implemented.

Based on the BESIII newest observation of the glueball-like particle X(2370) and our simulated results in $e^+e^-$ and pp collisions, we strongly suggest that the ALICE and LHCb collaborations, respectively, measure the  X(2370) production in pp collisions at the LHC energy in the mid-rapidity and forward-rapidity regions. It will be  
lighting up the determination of the X(2370) nature.

It is worth extending the investigation  to Pb-Pb collisions at $\sqrt{s_{\rm NN}}=$ 2.76 and 5.02 TeV. The ultra-relativistic heavy-ion collisions might be an ideal tool to produce glueball as the existence of quark-gluon plasma (QGP) provides a large amount of thermal gluons \cite{glueball_mass_predict_1}.  In this gluon-rich environment, the gluons can form glueballs. With the QGP cooling further and freezing-out into hadronic matter, these glueballs might decay into light hadrons, which will give rise to signatures of their existence.

\begin{acknowledgments}
We would like to thank San Jin, Yan-Ping Huang, Zhi-Qing Liu and Li-Lin Zhu for the valuable 
discussions. This work is supported by the National Natural Science
Foundation of China under grant Nos. 11447024, 11505108 and 12375135, and by the 111 project of the
foreign expert bureau of China. Y.L.Y. acknowledges the financial support
from Key Laboratory of Quark and Lepton Physics in Central
China Normal University under grant No. QLPL201805 and the Continuous Basic
Scientific Research Project (No, WDJC-2019-13). W.C.Z. is supported
by the Natural Science Basic Research Plan in Shaanxi Province of China
(No. 2023-JCYB-012). H.Z. acknowledges the financial support from
Key Laboratory of Quark and Lepton Physics in Central China Normal University
under grant No. QLPL2024P01.

\end{acknowledgments}

\end{document}